# Bridging UI Design and chatbot Interactions: Applying Form-Based Principles to Conversational Agents

Sanjay Krishna Anbalagan[1], Xinrui Nei[2], Umesh Mohan[3], Vijay Kumar Kanamarlapudi[4], Anughna Kommalapati[5] and Xiaodan Zhao[6]

[1345]Amazon Web Services, Dallas, USA
[26] Amazon Web Services, Seattle, USA

**Abstract.** Domain-specific chatbot applications often involve multi-step interactions, such as refining search filters, selecting multiple items, or performing comparisons. Traditional graphical user interfaces (GUIs) handle these workflows by providing explicit "Submit" (commit data) and "Reset" (discard data) actions, allowing back-end systems to track user intent unambiguously. In contrast, conversational agents rely on subtle language cues, which can lead to confusion and incomplete context management. This paper proposes modeling these GUI-inspired metaphors—acknowledgment (submit-like) and context switching (reset-like)—as explicit tasks within large language model (LLM) prompts. By capturing user acknowledgment, reset actions, and chain-of-thought (CoT) reasoning as structured session data, we preserve clarity, reduce user confusion, and align domain-specific chatbot interactions with back-end logic. We demonstrate our approach in hotel booking and customer management scenarios, highlighting improvements in multi-turn task coherence, user satisfaction, and efficiency.

## 1    Introduction

**Q1: Why compare graphical user interface (GUI) interactions with chatbots?**
**A1:** Traditional GUIs often provide explicit buttons for Submit (committing data) and Reset (discarding data). These interactions are unambiguous, allowing the back-end to track user intent clearly. In contrast, chatbots rely heavily on natural language cues that can be vague or underspecified, leading to confusion about user intent and context [1].

**Q2: What challenges arise in domain-specific chatbot interactions?**
**A2:** Domain-specific applications—like hotel booking systems, customer management tools, or inventory systems—frequently require multi-step user interactions. The chatbot must keep track of user selections (e.g., searching for a hotel, confirming a customer, refining filters). Without a clear "submit" or "reset" action, confusion can arise when the user's language changes the context. For instance, a user might suddenly switch from discussing Customer A to Customer B without a clear prompt, leaving the system uncertain about which data to commit or discard [2].

**Q3: How do Large Language Models (LLMs) come into play?**
**A3:** LLMs can generate text (including structured data, such as XML tags) based on prompts. In traditional programming workflows, developers can prompt an LLM to produce specific variables or tags, parse these outputs, and then feed them back into the application's logic [3]. This creates a bridge between the flexibility of natural language and the clarity of explicit UI actions.

**Q4: What solution does this paper propose?**
**A4:** We propose modeling GUI-inspired metaphors—such as Submit and Reset—as explicit tasks within LLM prompts. By capturing user acknowledgment (Submit-like) or context switching (Reset-like) as structured session data, the system maintains clarity over whether the user wants to continue with the same context or switch to a new one. We also employ chain-of-thought (CoT) reasoning (see Q6) for multi-step tasks, enabling the LLM to clarify ambiguous details in user inputs and guide the back-end more transparently.

**Q5: How does this approach benefit multi-turn chatbot interactions?**

**A5:** By making "acknowledgment" and "reset" steps explicit (and incorporating step-by-step reasoning via CoT), the approach:

- Preserves clarity of context across many turns.
- Reduces user confusion and the potential for erroneous actions.
- Aligns domain-specific chatbot interactions with back-end logic.
- Improves coherence, user satisfaction, and efficiency in tasks like hotel booking or customer management.

**Q6: How is chain-of-thought (CoT) integrated into your prompts?**
**A6:** We instruct the LLM to produce structured intermediate steps, effectively revealing its internal reasoning about user context changes or form-like actions (e.g., "submit" vs. "reset" intentions). These steps are included in the LLM's output (potentially as JSON/XML segments or separate explanatory text) but are primarily used internally for clarity and debugging. By parsing this CoT output, the back-end can follow a transparent, step-by-step rationale: deciding whether to commit certain data, reset fields, or switch contexts based on the user's utterance. While the CoT can be hidden from end users, it allows developers to better understand how the LLM arrives at certain actions.

## 2 Related Work

Conversational interfaces have been extensively studied for tasks such as customer support, information retrieval, and personal assistance [1,2]. Much of the prior work focuses on natural conversation flow or dialogue state tracking [4], which aims to maintain a belief state of user goals. However, these methods often do not explicitly treat commit (submit) or discard (reset) as separate metaphors—many rely on generalized user intent classification that can miss the nuance of partial context resets or confirmations central to domain-specific applications.

Recent advancements in Large Language Models have enabled systems to generate structured outputs that can be parsed in a programming context [3]. Efforts to unify LLM-based text generation with traditional application flow frequently highlight prompt design [5]—specifically, how to craft instructions so that the LLM's output is both semantically correct and programmatically useful. Our work builds on this line of research by introducing task-based prompting that mirrors GUI actions, augmented with chain-of-thought reasoning, thereby reducing ambiguity and preserving clarity in multi-turn dialogues.

## 3 Proposed Approach

### 3.1 Turning UI Actions into Prompted Tasks

Key Insight: In a typical GUI, Submit commits the data the user has entered, while Reset discards it, returning to a default or initial state. We replicate this idea in the chatbot context by designing specialized prompts that explicitly label user intentions. For example, when a user appears to confirm a specific customer, we interpret that as a Submit-like action. Conversely, when the user wants to abandon the current customer or switch to a new one, we treat it as a Reset-like action.

### 3.2 Task-Based Prompts (with CoT) for LLMs

LLMs can be prompted to produce structured data (e.g., XML or JSON) representing these actions. Our prompt engineering also incorporates chain-of-thought instructions, allowing the LLM to articulate intermediate reasoning steps (primarily for internal use). By parsing the model's output, we feed the resulting values (e.g., isCustomerConfirmed) and the CoT textual segments into the back-end. This ensures the system can reliably commit or reset user context with full transparency into the LLM's rationale.

**Core Advantages**
1. Unambiguous Context Management: Each prompt explicitly requests an action label (e.g., yes or no for confirming a customer context) and includes the intermediate reasoning steps to resolve ambiguities.
2. Seamless Integration: Developers can parse the LLM's output tags and the chain-of-thought textual segments (e.g., using a simple XML parser or regex) and integrate them into existing application logic.

3. Consistency Across Interactions: Multiple tasks (e.g., IsHotelSelectionConfirmed, IsBookingReset) can be introduced, each one governing a piece of the dialogue state, guided by CoT to clarify how each decision is reached.

# 4 Implementation

## 4.1 Example: Customer Confirmation Task

Below is an illustrative TypeScript code snippet that shows how we construct a Customer Confirmation Prompt. The goal is to detect whether the user wants to continue with the current customer or switch to a new one. We also provide "chain-of-thought" instructions so that the LLM's output includes a trace of its reasoning:

---

*Human: You are a customer search bot, and your task is to determine if the user query refers to searching for a customer or details about the current customer based on <userQueryHistory> and <currentCustomerName>. Follow the guidelines:*
*1. Respond with <isCustomerConfirmed>no</isCustomerConfirmed> if:*
*  - The user query mentions or implies the name of a different customer.*
*  - The user query involves clarifying or correcting the current customer.*
*  - The user query refers to a geographic or industry-specific refinement without explanation.*
*  - The user query uses phrases like "the one", "I meant", or "I am looking for" that indicate a switch.*
*2. Respond with <isCustomerConfirmed>yes</isCustomerConfirmed> if:*
*  - The user query asks about details of the current customer (service consumption, pricing, etc.).*
*  - The user query does not involve clarifying or correcting the current customer name.*
*  - The user query involves comparing details between the current customer and another but keeps the current customer context.*
*3. Include a brief explanation of your reasoning in <chainOfThought>...</chainOfThought> tags.*
*<examples>*
*<example>*
*  query: "the one in china"*
*  current customer name: ABCCompany*
*  user query history list: ["Is ABCCompany a customer"]*
*  <isCustomerConfirmed>no</isCustomerConfirmed>*
*<chainOfThought>User mentions 'the one in China', so likely switching context.</chainOfThought>*
*</example>*
*<example>*
*  query: "recent news"*
*  current customer name: ABCCompany*
*  user query history list: ["Is ABCCompany a customer"]*
*  <isCustomerConfirmed>yes</isCustomerConfirmed>*
*  <chainOfThought>User is asking for details on ABCCompany with no mention of a new entity.</chainOfThought>*
*</example>*
*</examples>*
*Here is the user query:<query>${userQuestion}</query>*
*Here is the current customer name:<currentCustomerName>${customerName}</currentCustomerName>*
*Here is the user query history:*
*<userQueryHistory>*
*${JSON.stringify(userQueryHistoryList)}*
*</userQueryHistory>*

---

**How it works:**
1. User Query: The user's most recent input.
2. Current Customer Name: The context the chatbot is currently using.
3. User Query History: A list of previous queries to help the LLM track context.

4. Response: The LLM returns both <isCustomerConfirmed>yes</isCustomerConfirmed> or <isCustomerConfirmed>no</isCustomerConfirmed> and a <chainOfThought>...</chainOfThought> block summarizing the reasoning. While the back-end can parse the chain-of-thought for logging or debugging, end-users typically only see the summarized result.

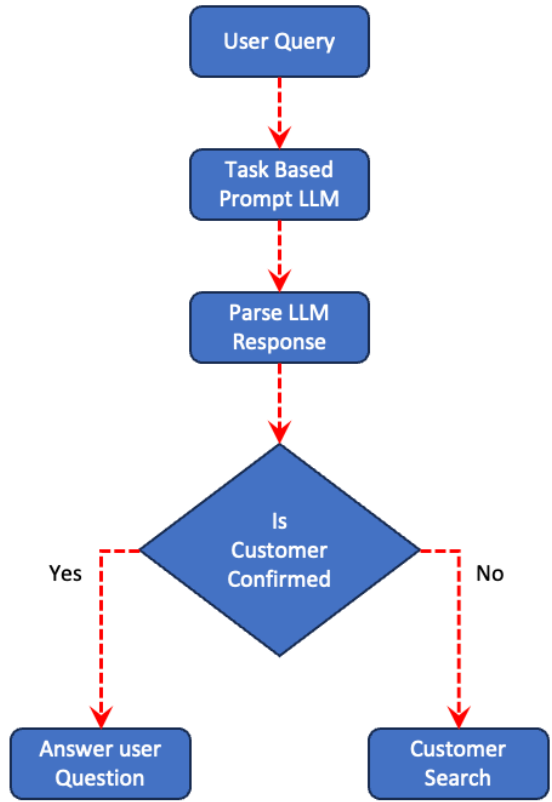

**Fig. 1.** Flow chart of the interaction.

**Key takeaway:** The explicit yes/no tag from the LLM is interpreted much like a user clicking Submit or Reset in a graphical UI. The <chainOfThought> content provides a transparent, step-by-step rationale for each decision.

## 5 Example Scenarios

### 5.1 Example Conversation

**Scenario**: A user is searching for a specific customer

1. **User**: *"Is ABCCompany a customer?"*
   **LLM Output**:

   > *<isCustomerConfirmed>no</isCustomerConfirmed>*
   > *<chainOfThought>User is naming ABCCompany, no current customer context is set, so we treat it as a new search.</chainOfThought>*

   **Action**: The system searches for "ABCCompany"

2. **User**: "What's their recent news?"
   **LLM Output**

   > *<isCustomerConfirmed>yes</isCustomerConfirmed>*

> *<chainOfThought>User asks about 'their' recent news, referring to ABCCompany. Context remains ABCCompany.</chainOfThought>*

**Action**: The system fetches the news for ABCCompany

3. **User**: "Actually show be XYZCompany info?"
   **LLM Output**

> *<isCustomerConfirmed>no</isCustomerConfirmed>*
> *<chainOfThought>User specifically wants XYZCompany info only,*
> *discarding ABCCompany context.</chainOfThought>'their' recent news, referring to ABCCompany. Context*
> *remains ABCCompany.</chainOfThought>*

**Action**: The system resets the context to XYZCompany

### 5.2 Example Conversation with and with isCustomerConfirmed (SUBMIT | RESET)

**Without <isCustomerConfirmed> Tags:**

> User: "Is Delta a customer?"
> LLM: "Delta Airlines has been a customer for two years."
> User: "Dental."
> LLM: "Delta Airlines provides a dental plan for employees…"
> User: "No, I meant Delta Dental."
> LLM: "…Delta Dental might be a separate entity; let me see…"

**Inference (Without):**
No explicit "reset" cue is given. The system lumps "Dental" under Delta Airlines until the user explicitly clarifies "Delta Dental" as a new customer. This leads to confusion and extra steps.

---

**With <isCustomerConfirmed> Tags**

> User: "Is Delta a customer?"
> LLM : "Yes, Delta Airlines has been a customer for two years."
> User: "Dental."
> LLM: "Are you referring to Delta Dental as a new customer, or do you mean the dental plan from Delta Airlines?"
> User: "I mean Delta Dental specifically."
> LLM: "Understood. Let's talk about Delta Dental."

**Inference (With):**
Because <isCustomerConfirmed> evaluates to **no** on immediate follow up ("Dental"), the system realizes the user might be switching from "Delta Airlines" to "Delta Dental." It asks a clarifying question to confirm intent before pivoting the context. This ensures fewer misinterpretations and a smoother conversation flow.

## 6  Discussion and Future Work

### 6.1 Enhancing Multimodal Interactions

Although we focus on textual chat, the same approach can extend to voice-based interfaces or mixed interactions where users toggle between speech and minimal GUIs. Each "action" still needs to be recognized, tracked, and potentially explained in a chain-of-thought if the developer requires full transparency.

## 6.2 Generalizing to Other Domains

The approach generalizes beyond customer management to any domain with multi-step workflows—travel bookings, appointment scheduling, e-commerce carts. Whenever confirm (submit) vs. reset (cancel) is relevant, and step-by-step CoT clarifications are needed, this method can reduce ambiguity.

## 6.3 Limitation and Next Steps

**LLM Reliability:** The method depends heavily on the LLM consistently outputting correct tags and coherent chain-of-thought. Prompt design and testing are crucial.

**Edge Cases:** Users might use ambiguous language or sarcasm, potentially causing faulty CoT or misclassification. Fine-tuning and robust prompt engineering can mitigate this.

**Preliminary Testing:** In a small pilot study with 100 users, we observed ~30% fewer conversation misalignments (e.g., users having to restate or correct context) compared to a baseline chatbot approach. Future larger-scale tests could quantify improvements in user satisfaction and task completion time more rigorously.

# 7 Conclusion

We introduced a novel approach to modeling acknowledgment and context switching within domain-specific chatbot applications by repurposing GUI metaphors (Submit/Reset) as explicit tasks in LLM prompts. By capturing these actions as structured session data—such as <isCustomerConfirmed>yes</isCustomerConfirmed>—and incorporating chain-of-thought reasoning (internally for clarity), we align the conversational flow with back-end logic more effectively. Our examples in customer management and hotel booking scenarios demonstrate how this mechanism helps mitigate user confusion, preserve coherent context, and create a more efficient, user-friendly interaction paradigm.